# Land Use and Land Cover Classification using a Human Group based Particle Swarm Optimization Algorithm with a LSTM classifier on hybrid-pre-processing Remote Sensing Images

R. Ganesh Babu[1], K. Uma Maheswari[2], C. Zarro[3], B. D. Parameshachari[4], and S. L. Ullo[5*]

[1,2] Department of Electronics and Communication Engineering, SRM TRP Engineering College, Tiruchirappalli, TN, India; [1]ganeshbaburajendran@gmail.com, [2]umaragsug@gmail.com

[3,5] Engineering Department, University of Sannio, Italy; [3]czarro@unisannio.it, [5]ullo@unisannio.it

[4] Department of Telecommunication Engineering, GSSS Institute of Engineering and Technology for Women, Mysuru, Karnataka, India; parameshbkit@gmail.com

* Correspondence: ullo@unisannio.it

**Abstract:** Land use and land cover (LULC) classification using remote sensing imagery plays a vital role in many environment modeling and land use inventories. In this study, a hybrid feature optimization algorithm along with a deep learning classifier is proposed to improve the performance of LULC classification, helping to predict wildlife habitat, deteriorating environmental quality, haphazard, etc. LULC classification is assessed using Sat 4, Sat 6 and Eurosat datasets. After the selection of remote sensing images, normalization and histogram equalization methods are used to improve the quality of the images. Then, a hybrid optimization is accomplished by using the Local Gabor Binary Pattern Histogram Sequence (LGBPHS), the Histogram of Oriented Gradient (HOG) and Haralick texture features, for the feature extraction from the selected images. The benefits of this hybrid optimization are a high discriminative power and invariance to color and grayscale images. Next, a Human Group based Particle Swarm Optimization (PSO) algorithm is applied to select the optimal features, whose benefits are fast convergence rate and easy to implement. After selecting the optimal feature values, a Long Short Term Memory (LSTM) network is utilized to classify the LULC classes. Experimental results showed that the Human Group based PSO algorithm with a LSTM classifier effectively well differentiates the land use and land cover classes in terms of classification accuracy, recall and precision. An improvement of 2.56% in accuracy is achieved compared to the existing models GoogleNet, VGG, AlexNet, ConvNet, when the proposed method is applied.

## 1. Introduction

In recent years, land use and land cover classification using remote sensing imagery plays an important role in many applications like land use planning (growth trends, suburban sprawl, policy regulations and incentives), agricultural practice (conservation easements, riparian zone buffers, cropping patterns and nutrient management), forest management (harvesting, health, resource-inventory, reforestation and stand-quality) and biological resource (fragmentation, habitat quality and wetlands) [1-3]. In addition, the land use and land cover assessments are very necessary in sustaining, monitoring and planning the usage of natural resources [4-5]. The land use and land cover classification have a direct impact on atmospheric, soil erosion and water where it is indirectly connected to global environmental problems [6-7]. The remote sensing imagery delivers up-to date and large scale information on surface condition. Present remote sensing images have two major concerns, those are noise associated with the image and maintaining the large volume of data [8-9]. Several methodologies have been developed by the researchers to address those issues of land use and land cover classification. Some of the methodologies are adaptive reflectance fusion model, maximum likelihood classifier, decision tree, deep convolutional neural network, Deep Neural Network (DNN), etc. [10-14]. The conventional techniques used in land use and land cover classification are extremely affected by the environmental changes like destruction of essential wetlands, uncontrolled urban development, haphazard, loss of prime agricultural lands, deteriorating environmental quality, etc. and also due to the factors like cloud cover and regional fog error.

The study presented in this article has proposed a new algorithm, a Human Group based Particle Swarm Optimization (PSO) algorithm, with a LSTM classifier, to address the above discussed issues and improve the land use and land cover classification, in agriculture and urban environment, especially in those cases not related to human habitants. Initially, the remote sensing images are retrieved from Sat 4, Sat 6 and Eurosat datasets [15-17]. After the selection of remote sensing images, normalization and histogram equalization methods are applied to improve the visual quality of the objects. The undertaken pre-processing techniques effectively improve the contrast of the images and enhance the edges in each region of an image. After normalizing the images, feature extraction is carried out by Local Gabor Binary Pattern Histogram Sequence (LGBPHS), Histogram of Oriented Gradient (HOG) and Haralick texture features [18-21]. The LGBPHS is utilized as a two-dimension spatial image gradient measurement to emphasize the high spatial frequency regions based on the image edges. In Addition, it is utilized to identify the absolute gradient scales at each point in a remote sensing image. Next, HOG and Haralick texture features are applied to extract the texture and color feature vectors from the image pixels. As, the HOG feature descriptor operates on local cells, so it is invariant to photometric and geometric transformations, that helps in attaining better classification. The textural properties are calculated by Haralick texture features in order to understand the edge details about the image content. Then, the Human Group based PSO algorithm is used to select the optimal feature vectors that significantly reduces the "curse of dimensionality" issue. The obtained optimal feature vectors are given as the input to the LSTM classifier to establish the land use and land cover classes. In the result section, the performance of the proposed Human Group based PSO with LSTM is evaluated in terms of recall, precision and classification accuracy and compared with other existing models: GoogleNet, VGG, AlexNet and ConvNet.

This research paper is organized as follows. Section II presents several existing research papers on the topic "land use and land cover classification", when HOG, LGBPHS, Machine Learning (ML), Object Based Image Analysis (OBIA), Bag-Of-Visual Words (BOVW) and Scale-Invariant Feature Transform (SIFT) methodologies are used. In Section III, the proposed model is briefly explained with mathematical expressions. Experimental analysis of the proposed model is then presented in the Section IV. The conclusions of this study are drawn in Section V.

**2. Literature survey**

Xiao et al. [22] developed a new rotationally invariant feature descriptor to identify cars and aircraft in the remote-sensing images. The rotationally invariant HOG feature descriptor used elliptic Fourier transform, orientation normalization and feature space mapping to achieve better performance in object detection from remote sensing images. Rahmani and Behrad [23] developed a new model for ship detection in the remote sensing images. Initially, the collected images were divided into overlapping blocks and then the LGBPHS feature descriptor was used to extract the features from the images. Support Vector Machine (SVM) and Artificial Neural Network (ANN) were used for classification after feature extraction. However, SVM supports only binary classification, which is adaptable for multiclass classification.

Kadavi and Lee [24] used SVM and ANN classifiers to evaluate the multi-spectral data from mount Fourpeaked, mount Kanaga, mount Augustine and mount Pavlof. In this study, a Landsat-8 imagery dataset was used to evaluate the efficiency and effectiveness of the developed model. The Landsat-8 imagery dataset contains four land cover classes vegetation, snow, water bodies and outcrops (sand, volcanic rock, etc.). Simulation results showed that the SVM classifier attained better performance in land use and land cover classification compared to ANN classifier. For mount Kanaga, the SVM classifier achieved maximum classification accuracy, which was 9.1% superior to ANN classifier. The developed model was only suitable for minimum class classification not for maximum class classification and the developed model showed poor performance in some conditions like cloud cover and regional fog error. Pencue-Fierro et al. [25] presented a new hybrid framework for multi-region, multi-sensor and multi-temporal satellite image classification. In this study, land cover classification was assessed for the Cauca river region, located in the south-west part of Colombia. After image collection, Coordination of Information on the Environment (CORINE) land cover approach was used for extracting the feature vectors from the input image. Next, the extracted features were given as the input to a supervised classifier SVM to classify the land cover classes like urban-area, paramo, snow, clouds, bare soil, grass-land, planted forest, permanent-crops, natural forest, water-bodies and transitory crops. However, the computational complexity was higher in the developed hybrid framework compared to the other methods.

Phiri et al. [26] evaluated moderate resolution atmospheric transmission, atmospheric correction, cosine topographic correction and dark object subtraction on a heterogeneous landscape in Zambia. In this study, Landsat OLI-8 with 30 and 15 m spatial resolution images were tested using a combination of random forest classifier [27] and Object Based Image Analysis (OBIA) [28]. The developed method significantly improved land cover classification along with topographic corrections and pan-sharpening atmosphere. The developed framework (random forest and OBIA) effectively classified eight land-cover class water bodies, grassland, secondary-forests, dry-agriculture, primary forests, irrigated crops, settlements and plantation-forests. This study did not concentrate on the feature extraction that may degrade the performance of land cover classification.

Zhao et al. [29] implemented a new framework for land use classification using UCMerced land-use dataset and simulated dataset. After collecting the satellite data, Bag-Of-Visual Words (BOVW) and Scale-Invariant Feature Transform (SIFT) [30] methodologies were used for extracting the feature vectors from the collected data. In addition, concentric circle based spatial rotation invariant representation was used to describe the spatial information of data. A concentric circle structured multi scale BOVW was used for land use classification. The performance of the developed method was analyzed in terms of average classification accuracy. However, the developed method fails to achieve better land use classification in the large datasets due to "curse of dimensionality" issue.

Nogueira et al. [31] used Convolutional Neural Network (CNN) in different scenarios like feature extraction, fine tuning and full training for land cover classification. The developed model's performance was investigated on three remote sensing datasets Brazilian coffee scene, UCMerced land use and remote sensing 19. The results indicated that the developed model attained better performance in land cover classification compared to the existing algorithms. Helber et al. [15] developed a new patch based land use and land cover classification technique using Eurosat dataset. The undertaken dataset had 13 spectral bands and 10 classes with a total of 27,000 geo-referenced and labeled images. This study explained how CNN was used to detect the land use and land cover changes that helped in improving the geographical maps. However, using middle and lower level descriptors, the CNN model leads to poor classification performance because it supports only higher level descriptors. Unnikrishnan et al. [16] developed a new deep learning model for three different networks VGG, AlexNet and ConvNet where Sat-4 and Sat-6 datasets were used to analyze the performance of the developed model. This model includes information on Red and Near Infrared bands, with reduced number of filters, which were tested and trained to classify the images into different classes. The developed model was compared with other networks in terms of trainable parameters, recall and classification accuracy. Papadomanolaki et al. [17] developed a deep learning model based on CNN for precise land cover classification. The performance of the developed model was compared with the existing networks AlexNet-small, AlexNet and VGG in terms of accuracy and precision on Sat-4 and Sat-6 datasets. However, the CNN has two major concerns computationally high cost and more data required to achieve precise classification.

Jayanth, et al. [52] developed an elephant herding algorithm to classify land use and land cover regions from high spatial resolution multi-spectral images. The developed elephant herding algorithm achieved high classification accuracy compared to SVM classifier. Experimental results showed that the elephant herding algorithm attained better performance in land use and land cover classification on both Arsikere taluk and NITK campus datasets. Extensive experiments showed that the elephant herding algorithm misclassified the dense coconut tree class that is considered as a major concern in this literature study. Bhosle and Musande [53] developed CNN model for crop classification, and land use and land cover classification in the hyperspectral remote sensing images. In this study, the CNN model works well on un-structured data, where it automatically extracts features for detection and classification of crops. The extensive experiment showed that the CNN model achieved effective performance on Indian Pines dataset. However, the CNN model does not encode the orientation and position of objects (crop types) and has lack of ability to be spatially invariant to the input data.

As already highlighted, a Human Group based PSO with a LSTM classifier is proposed in this study to address the above discussed issues and improve the land use and land cover classification, as it will be shown and discussed in the last sections.

## 3. Method

Land use and land cover which are present on the surface of the earth are known as the elements or features, including the natural wilderness or environments like settlements, semi natural habitats like pastures, managed woods, arable fields, etc. [32-33]. The land cover features refer to the changes in biodiversity, erosion, modification and conversion of vegetation, soil-quality, sedimentation and land productivity [34-35]. The awareness about land use and land cover classification is very essential to address the concerns of destruction of central wetlands, wildlife habitat, deteriorating environmental quality, haphazard, loss of prime agricultural lands and uncontrolled development [36]. At this end, a Human Group based PSO with a LSTM classifier is proposed, where the complete workflow is graphically presented in figure 1, highlighting also the pre-processing steps with the optimization of feature extraction included.

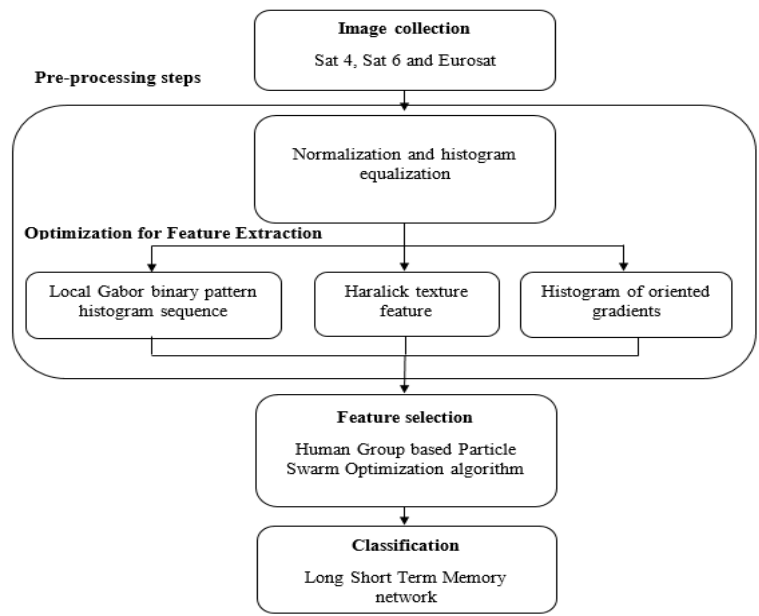

**Figure 1.** Work flow of the complete Human Group based PSO with LSTM, and the pre-processing steps including the feature extraction.

*3.1. Image collection*

In this study, Sat 4, Sat 6 and Eurosat databases are utilized for experimental analysis to differentiate the things that are not related to human habitats in both urban and agricultural environments. Sat 4 database comprises of 500,000 airborne images with four broad land cover classes like tree, barren land, grassland and a class with all land cover classes, except, tree, barren land and grassland [37]. The size of each remote sensing image in Sat 4 database is $28 \times 28$. Sat 6 database comprises of 40,500 airborne images with the size of $28 \times 28$ and it contains six land cover classes grassland, water bodies, buildings, barren land, roads and trees [16-17]. The sample image of Sat 4 and Sat 6 dataset is presented in figure 2.

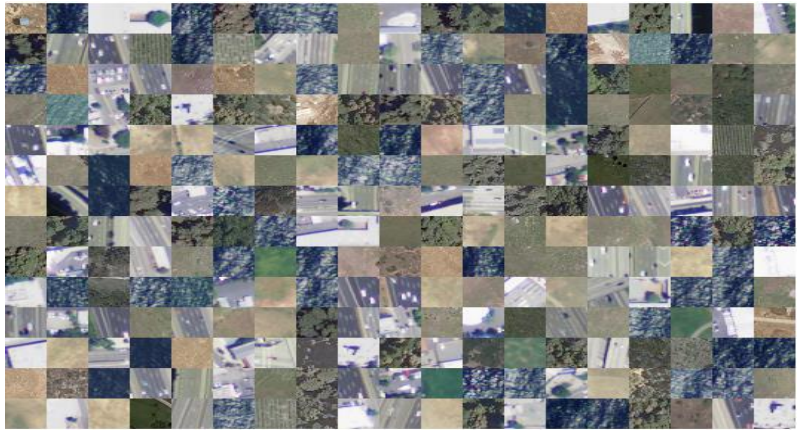

**Figure 2.** Sample image of Sat 4 and Sat 6 databases [16-17]

In Eurosat database, the satellite images have been captured from European cities, which distributed in over thirty-four countries. A database is generated with 27,000 labeled and geo referenced image patches, where the size of the image patch is $64 \times 64$. The Eurosat database includes 10 different classes, where each class contains 2,000-3,000 images. The land use and land cover classes in this database are permanent crop, annual crop, pastures, river, sea &lake, forest, herbaceous vegetation, industrial building, highway and residential building [15]. In addition, Eurosat images includes 13 bands like aerosols, blue, green, red, red edge 1, red edge 2, red edge 3, near infrared, red edge 4, water vapor, cirrus, shortwave infrared 1 and shortwave infrared 1. A sample image of Eurosat database is presented in figure 3.

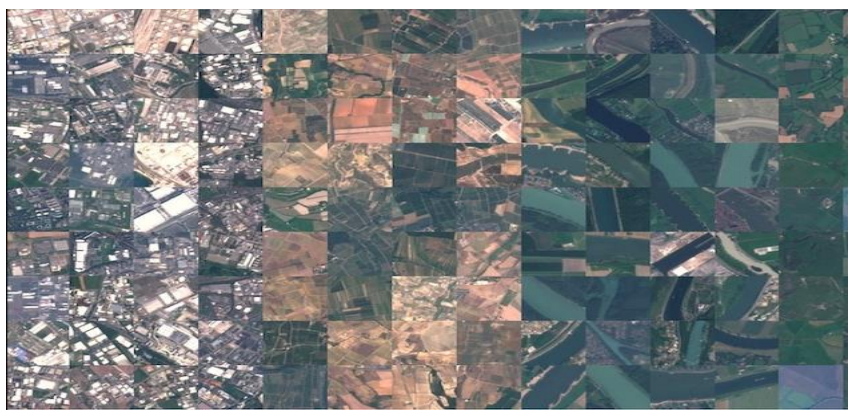

**Figure 3.** Sample image of Eurosat database [15]

*3.2. Image pre-processing*

After collecting the satellite images, normalization and histogram equalization methods are undertaken to improve the quality of the images. Image normalization is also called as contrast stretching that changes the range of pixel values which helps in improving the visual quality of the collected satellite images. General formulas of normalization techniques are presented in equation (1).

$$I\_out = (I\_in - Min)\frac{newMax - newMin}{Max - Min} + newMin \qquad (1)$$

where, original satellite image is indicated as $I\_in$, minimum and maximum intensity values are represented as $Min$ and $Max$ respectively, which ranges from 0 to 255, the image after the min-max normalization is indicated with $I\_out$, and the new minimum and maximum values are indicated with $NewMin$ and $newMax$. The common case of a min-max normalization to a new image ranging from 0 to 1, results into the well-known simplified version of the equation (1). Then, the histogram equalization technique is used to adjust the contrast of the images using the histogram values. In the image enhancement, histogram equalization is the best technique that delivers better image quality without losing the image information like edges, image patches and points [38-39].

*3.3. Feature extraction*

After normalization and histogram equalization of the collected satellite images, feature extraction is carried out by using a hybrid optimization procedure, based on the joint use of HOG, LGBPHS and Haralick texture features, namely correlation, contrast, energy, homogeneity, inverse diverse moment, entropy and angular second moment, to extract the feature vectors from the images.

**HOG**: In the satellite image, the HOG feature descriptor significantly captures the gradient and edge structure of the objects. Though, the HOG feature descriptor operates in the localized cells, which upholds invariance to photometric and geometric transformations except object orientation. This action helps in finding the changes appears in the large spatial regions. Here, a simple gradient operator $K$ is applied to determine the gradient value. The gradient of the image is given by equation (2), where $x, y$ represents a generic point in the image and the image frames are denoted as $u$.

$$L_x = K * u(x,y) \text{ and } L_y = K^T * u(x,y) \quad L_x \qquad (2)$$

The magnitude of the gradients and edge orientation of the point $x, y$ is calculated by following the respective conditions (Eq.3 and Eq.4),

$$L(x,y) = \sqrt{L_x(x,y)^2 L_y(x,y)^2} \qquad (3)$$

$$\theta(x,y) = \tan^{-1} L_y(x,y) / L_x(x,y) \qquad (4)$$

For improving the invariance in illumination and noise, a normalization process is performed after the calculation of histogram values. The normalization is helpful for contrast and measurement of local histogram. In HOG four different normalizations are used such as L2-norm, L2-Hys, L1-Sqrt and L1-norm. Among these normalizations, L2-norm gives better performance in object detection. The blocks of normalization in HOG is given by Eq. (5),

$$L_{2-norm} : f = \frac{q}{\sqrt{\|q\|_2^2 + e^2}} \qquad (5)$$

Where, $e$ is the small positive value, only when an empty cell is taken into account, $J$ is a feature extracted value, $q$ is the non-normalized vector in histogram blocks, and $\|q\|_2^2$ $\|q\|_2^2$ represents the 2-norm of HOG normalization.

**LGBPHS:** Initially, the pre-processed satellite images are transformed to obtain multiple Gabor Magnitude Pictures (GMP) using multi-orientation and multi scale Gabor filters. Then, each GMP is converted into Local GMP (LGMP) that is further categorized into non-overlapping rectangular regions with specific histogram and size [49]. The LGMP histogram of all the LGMO maps is combined to form final histogram sequences. Features extracted by LGBPH are robust to illumination variations, because the LGBPH features are invariant to monotonic gray-scale changes.

**Haralick texture features:** The Haralick features are $2^{nd}$ order statistics that reflect the overall average degree of correlation between the pixels in different aspects like contrast, energy, inverse difference moment, entropy, homogeneity, correlation and angular second moment [40]. The texture features are calculated from the texture information that are presented in the Grey-Level Co-occurrence Matrix (GLCM) [46]. In order to develop a number of spatial indices, Haralick uses the GLCM, because it contains the two neighboring pixels' relative frequencies by a distance on the image. Haralick developed the vast number of textural features with original 14 features that are described in [47], but only seven features are widely used due to its importance values for remote sensing images. Therefore, in this study, those seven commonly used features are considered as extracted features and this feature showed better performance in [48]. Although, Haralick texture feature effectively delivers information regarding the relative position of the neighborhood image pixels in the satellite images that helps in improving land use and land cover classification performance. A set of seven different GLCM indicators is described in the following equations (6 to 12);

$$Energy = \sum_{\gamma,\delta} \varphi(\gamma,\delta)^2 \qquad (6)$$

$$Entropy = \sum_{\gamma,\delta} \varphi(\gamma,\delta) \log_2 \varphi(\gamma,\delta), or\, 0\, if\, \varphi(\gamma,\delta) = 0 \qquad (7)$$

$$Correlation = \sum_{\gamma,\delta} \frac{(\gamma-\mu)(\delta-\mu)\varphi(\gamma,\delta)}{\sigma^2} \qquad (8)$$

$$Angular\, \sec ond\, Moment = \sum_{\gamma,\mu} \left( \varphi(\gamma,\delta) \right)^2 \qquad (9)$$

$$Inverse\, Difference\, Moment = \sum_{\gamma,\delta} \frac{1}{1+(\gamma-\delta)^2} \varphi(\gamma,\delta) \qquad (10)$$

$$Contrast = \sum_{z=0}^{N_g-1} z^2 \left\{ \sum_{\gamma=0}^{N_g} \varphi(\gamma,\delta) \right\} \qquad (11)$$

$$Homogenity = \sum_{\gamma,\delta} \frac{\varphi(\gamma,\delta)}{1+(\gamma-\delta)^2} \qquad (12)$$

Where, matrix cell index is depicted as $(\gamma,\delta)$, frequency value of the pair of index is represented as $\varphi(\gamma,\delta)$, mean and standard deviation of the row sums is illustrated as $\mu_t$ and $\sigma_t$ average of means weighted pixel is described $(\gamma,\delta)$ as $\mu = \sum_{\gamma,\delta} \gamma * \varphi(\gamma,\delta) = \sum_{\gamma,\delta} \delta * \varphi(\gamma,\delta)$, variance of means weighted pixel is defined as $\sigma = \sum_{\gamma,\delta} (\gamma-\mu)^2 * \varphi(\gamma,\delta) = \sum_{\gamma,\delta} (\delta-\mu)^2 * \varphi(\gamma,\delta)$, $N_g$, illustrates the total number of distinct gray levels in the images.

The variable importance analysis is carried out by the GLCM classification results, where the high importance of the variable is represented by high values of GLCM. From the experimental analysis in [48], the author P. Kupidura, proves that the Haralick's seven selected features have the highest significance among the 14 original features, by calculating the importance of these features. Therefore, the classification results proved that the Haralick features have higher resolutions, which are the best features rather than others for satellite image classification.

*3.4. Feature selection*

Feature selection is carried out by using the Human Group based PSO algorithm after extracting the feature vectors. Generally, PSO is a population based searching algorithm that mimics the behavior of birds. In order to generate new positions of every particle, equation (13) is used to update the velocity $v_i$ and position $p_i$ of the particles.

$$v_i(n+1) = w \times v_i(n) + r_1 \times c_1 \times \left( lb_i(n) - p_i(n) \right) + r_2 \times c_2 \times \left( gb_i(n) - p_i(n) \right)$$

$$p_i(n+1) = p_i(n) + v_i(n+1) \qquad (13)$$

where, $n$ is represented as the iteration, $r_1$ and $r_2$ are denoted as random numbers between [0, 1], $w$ is denoted as inertia weight, $b_i$ is indicated as the best position, $lb_i(n)$ is stated as local best position and $gb_i(n)$ is indicated as the global best position of the particle. In PSO, HGO algorithm is utilized initially to influence the particles and then the adaptive uniform mutation is utilized to improve the convergence rate and makes the implementation simple.

**Fitness function and encoding of particle:** Initially, HGO is used to transform discrete multi-label into continuous label. The undertaken algorithm finds the extracted feature vectors based on decision $d_i$, where the vectors of the particle's position are presented as $p_i(n) = (p_{i,1}, p_{i,2}, p_{i,D})$.

**Adaptive uniform mutation:** The adaptive uniform mutation is utilized to increase the ability of the feature selection algorithm in exploration. In this operator, a non-linear function $p_m$ $p_m$ is used to control the range and decision of the mutation on each particle $p_i$. At every iteration, $p_m$ is updated using equation (14).

$$p_m = 0.5 \times e^{\left(-10 \times \frac{n}{N}\right)} + 0 \qquad (14)$$

where, $N$ is indicated as maximum iteration, $n$ is denoted as number of iterations and the $p_m$ value tends to decrease when the number of iterations increases. The mutation randomly picks the $k$ elements from the particle, if the $p_m$ value is higher than the random number between [0,1]. Then, the mutation value of the elements within the search space is reinitialized, where $k$ is an integer value which is used for controlling the mutation range [41-42]. Mathematically, $k$ value is represented in equation (15), as:

$$k = \max\{1, |D \times p_m|\} \qquad (15)$$

The flow chart related to the Human Group based PSO algorithm is given in figure 4, with the further description of the steps below.

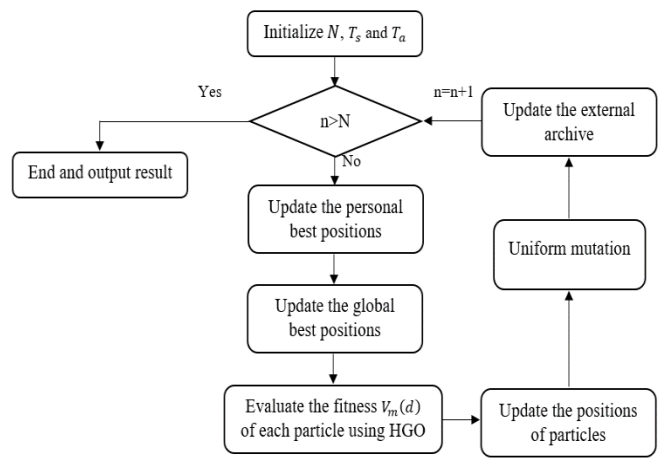

**Figure 4.** Flow chart of the Human Group based PSO algorithm

**Step 1:** Initialize the particles swarm, a) set the number of iterations $N$, swarm size $T_s$ and archive size $T_a$ b) initialize the particles location, c) estimate the objective of every particles, d) save non-dominated solution into the archive.

**Step 2:** Pareto domination relationship is used to update the personal best position of the particles. If new position $p_i(n+1)$ is better than old personal best position $lb_i(n)$, set $lb_i(n+1) = p_i(n+1)$, or else unchanged the personal best position of the particles, where $b_i$ is represented as best position and $lb_i(n)$ is presented as local best position.

**Step 3:** Based on the diversity of solution, select the global best position from the archive. At first, crowding distance value is calculated and then binary tournament is used to select the global best position of the particle $gb_i(n)$.

**Step 4:** Then, initialize the decision value $d_i$ based on $gb_i(n)$. Every decision $d_i$ of the feature vector $d$ is a binary value $d_i = \pm 1, i = 1, 2, \ldots T$, Every feature vector $d$ is related to the fitness value $V(d)$ that is considered as the weighted sum of $T$ stochastic contributions $W_j(d_j, d_1^j, \ldots, d_s^j)$. However, these contributions depend on the value of decision $d_j$ and other $S$ decisions $d_i^j, i = 1, 2, \ldots, S$. The fitness function is mathematically presented in equation (16):

$$V(d) = \frac{1}{T} \sum_{j=1}^{T} W_j(d_j, d_1^j, d_2^j, \ldots, d_s^j) \quad (16)$$

where, the integer index $S = 0, 1, 2, \ldots T - 1$ corresponds to the number of interacting decision values. The knowledge level of the $m^{th}$ $m^{th}$ member is determined by the parameter $P \in [0,1]$, which is the probability of each member that knows the contribution of the decision. On the basis of the knowledge level, every member $m$ computes own perceived fitness using equation (17):

$$V_m(d) = \frac{\sum_{j=1}^{T} \breve{d}_{mj} W_j(d_j, d_1^j, d_2^j, \ldots, d_s^j)}{\sum_{j=1}^{T} \breve{d}_{mj}} \quad (17)$$

where, $\breve{d}$ is denoted as the matrix, whose elements $\breve{d}_{mj}$ considers the value one with probability $P$ and 0) with probability $1 - P$.

**Step 5:** Based on the decision value $d_i$, equation (18) is used to update the velocity $v_i$ and position $p_i$ of the particles.

$$v_i(n+1) = w \times v_i(n) + r_1 \times V_m(d) \times (lb_i(n) - p_i(n)) + r_2 \times V_m(d) \times gb_i(n) - p_i(n))$$

$$p_i(n+1) = p_i(n) + v_i(n+1) \quad (18)$$

**Step 6:** Perform uniform mutation using the equations (14) and (15).
**Step 7:** Update the external archive using crowding distance methodology.
**Step 8:** Analyze the termination condition, if the proposed algorithm attains the maximum iteration stop the condition, or else return to step 2. Hence, the worst particles (feature vectors) are eliminated based on the fitness function $V_m(d)$ of HGO algorithm. In all three datasets, approximately 70%-80% of the feature vectors are selected from the total extracted features. After selecting the optimal features, classification is then carried out using the LSTM classifier. Table 1 states the extracted and the selected features after applying the Human Group based PSO algorithm. For instance; $38 \times 5000$, 38 represents number of images and 5000 denotes number of extracted features.

**Table 1.** Selected feature vectors after applying the feature selection algorithm

| Datasets | Extracted features | Selected features |
|----------|--------------------|--------------------|
| Sat 4    | $38 \times 5000$   | $38 \times 3671$   |
| Sat 6    | $50 \times 8700$   | $50 \times 6290$   |
| Eurosat  | $55 \times 9000$   | $55 \times 7098$   |

*3.5. Classification*

The LSTM classifier has the default behavior of remembering data information for a long time period [50-51]. In land use and land cover classification, a huge number of remote sensing images are needed to be proceeded for attaining better results. By considering this aspect, LSTM classifier is the best choice for land use and land cover classification [54-55]. Generally, the LSTM classifier is composed of a series of LSTM units, where the temporal quasi periodic features for extracting the long term and short term dependencies are stored. Hence, the structure of LSTM classifier is denoted in figure 5 and it includes 98 LSTM units that are graphically stated in figure 6.

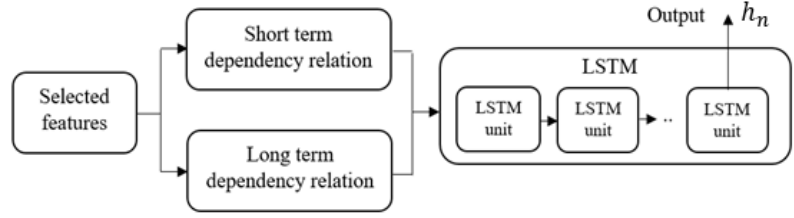

**Figure 5.** Structure of the LSTM classifier

The LSTM classifier contains an input gate $i_n$, a forget gate $f_n$, a cell $c_n$ and an output gate $o_n$, which are mathematically expressed in the equations (19-22).

$$i_n = \sigma(W_{ih}h_{n-1} + W_{ia}a_t + b_i) \quad (19)$$

$$f_n = \sigma(W_{fh}h_{n-1} + W_{fa}a_t + b_f) \quad (20)$$

$$c_n = f_n \times c_{n-1} + i_n \times \tanh(W_{ch}h_{n-1} + W_{ca}a_n + b_c) \quad (21)$$

$$o_n = \sigma(W_{oh}h_{n-1} + W_{oa}a_n + b_0) \quad (22)$$

where, $a_n = A[n,.] \in \mathbb{R}^F$ is represented as the quasi periodic feature in different frequency bands at the time step. Work coefficients are denoted as $W$ and $b$ and the hyperbolic tangent and sigmoid activation functions are indicated as $\tanh(.)$ and $\sigma(.)$. The output of the prior LSTM unit is stated as $h_{n-1}$ [43]. The output of the LSTM unit is mathematically denoted in equation (23).

$$h_n = o_n \times \tanh(c_n) \quad (23)$$

As shown in figure 5, $h_n$ contains the information of the prior time steps by $c_n$ and $o_n$. On the basis of dependency relation, the cell state $\{c_n | n = 1, 2, , N\}$ learns the memory information of the temporal quasi-periodic features for a long and short period of time during the training process. At last, the extracted features are denoted by the output of last LSTM unit $h_N$.

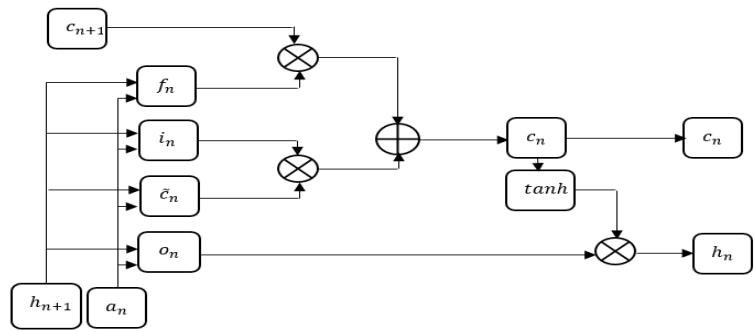

**Figure 6.** Graphical representation of the LSTM unit

## 4. Result and Discussion

The proposed model is simulated using MATLAB 2019 on a PC with 128 GB RAM, i9 Intel core processor, Windows 10 operating system (64-bit) and 3 TB hard disk. The performance of the proposed model is compared with a few benchmark models: GoogleNet [15], 2 band VGG [16], hyper parameter tuned VGG [16], 2 band AlexNet [16], hyper parameter tuned AlexNet [16], 2 band ConvNet [16], hyper parameter tuned ConvNet [16], AlexNet [17], ConvNet [17] and VGG [17], in order to find out its effectiveness. Specifically, the performance of the proposed model is evaluated on the selected datasets in terms of precision, recall and accuracy parameters. The mathematical expressions of accuracy, recall and precision are represented in the following equations (24-26).

$$Accuracy = \frac{TP+TN}{FN+TP+TN+FP} \times 100 \qquad (24)$$

$$\mathrm{Re}\,call = \frac{TP}{TP+FN} \times 100 \qquad (25)$$

$$\mathrm{Pr}\,ecision = \frac{TP}{TP+FP} \times 100 \qquad (26)$$

where, true negative is denoted as $TN$, false negative is represented as $FN$, true positive as $TP$ and false positive as $FP$.

*4.1. Quantitative investigation on Sat 4 database*

Sat 4 database is used to evaluate the performance of the proposed model to classify four land cover classes: tree, barren land, grass land and a class with all land cover classes, except tree, barren land and grass land. In this case, the performance evaluation is validated by using 500,000 satellite images with 70% of the data used for training and 30% for testing. Tables 3 and 4 represent the performance evaluation of the proposed model with different classifiers; DNN, Multi Support Vector Machine (MSVM) and LSTM in terms of classification accuracy, recall and precision. Two different case studies are considered in the experiments. The tables 2 and 3 point out that the LSTM classifier achieves better classification performance in land use and land cover classification on various classes by means of precision, recall and classification accuracy. Tables 4 and 5 illustrate that the LSTM classifier improves the accuracy in land use and land cover classification on various classes up to 1% compared to DNN and LSTM classifiers. Table 6 shows the performance of various neural networks with proposed LSTM for overall Sat 4 database in terms of classification accuracy. Compared to other classifiers, the LSTM has the ability to remember data information for a long period of time, where this behavior helps to attain better performance in land use and land cover classification. Performance analysis of the proposed model with different classifiers on Sat 4 dataset is represented in figure 7.

**Table 2.** Performance investigation of the proposed model with different classifiers on Sat 4 dataset by means of recall and precision

| Classes | Human group based PSO with DNN | | Human group based PSO with MSVM | | Human group based PSO with LSTM | |
|---|---|---|---|---|---|---|
| | Precision (%) | Recall (%) | Precision (%) | Recall (%) | Precision (%) | Recall (%) |
| Barren land | 99.07 | 99.12 | 99.65 | 99.79 | 99.90 | 99.98 |
| Trees | 99.54 | 99.65 | 99.67 | 99.65 | 99.98 | 99.97 |
| Grasslands | 99.60 | 99.87 | 99.43 | 99.80 | 99.97 | 99.95 |
| Others | 99.61 | 99.90 | 99.60 | 99.87 | 99.98 | 99.97 |
| Overall | 99.45 | 99.63 | 99.58 | 99.77 | 99.95 | 99.96 |

**Table 3.** Performance investigation of the proposed model with different classifiers on Sat 4 dataset by means of accuracy

| | Classification accuracy (%) | | |
|---|---|---|---|
| Classes | Human group based PSO with DNN | Human group based PSO with MSVM | Human group based PSO with LSTM |
| Barren land | 98.90 | 99.19 | 100 |
| Trees | 99.57 | 98.64 | 99.98 |
| Grasslands | 98.80 | 98.80 | 99.99 |
| Others | 99.15 | 98.76 | 100 |
| Overall | 99.10 | 98.84 | 99.99 |

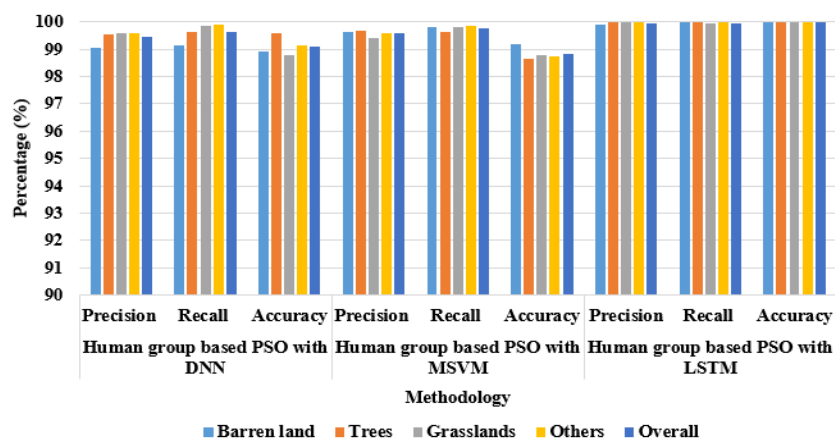

**Figure 7.** Graphical investigation of the proposed model with different classifiers on Sat 4 dataset

The tables 4 and 5 present the performance evaluation of the proposed model with different optimization techniques like PSO, HGO and human group based PSO, by pointing out that the LSTM classifier with Human Group based PSO achieves better performance in land use and land cover classification in terms of precision, classification accuracy and recall. The proposed model; human group based PSO with LSTM showed a maximum of 6.025% and a minimum of 1.66% improvement in land use and land cover classification, if compared to LSTM, PSO with LSTM and HGO with LSTM. The figure 8 shows the performance analysis of the proposed model when different optimization techniques are applied on Sat 4 dataset.

**Table 4.** Performance investigation of the proposed model with different optimization techniques on Sat 4 dataset by means of recall and precision

| Classes | LSTM | | PSO with LSTM | | HGO with LSTM | | Human group based PSO with LSTM | |
|---|---|---|---|---|---|---|---|---|
| | Precision (%) | Recall (%) | Precision (%) | Recall (%) | Precision (%) | Recall (%) | Precision (%) | Recall (%) |
| Barren land | 94.65 | 94.92 | 98.74 | 98.16 | 98.62 | 98.70 | 99.90 | 99.98 |
| Trees | 93 | 95 | 96.56 | 97.55 | 98.69 | 98.73 | 99.98 | 99.97 |
| Grasslands | 94.90 | 96.10 | 97.88 | 97.82 | 98.93 | 98.98 | 99.97 | 99.95 |
| Others | 93.70 | 94 | 98.63 | 98.87 | 98.69 | 98.92 | 99.98 | 99.97 |
| Overall | 94.06 | 95.005 | 97.95 | 98.10 | 98.65 | 98.83 | 99.95 | 99.96 |

**Table 5.** Performance investigation of the proposed model with different optimization techniques on Sat 4 dataset by means of classification accuracy

| | Classification accuracy (%) | | | |
|---|---|---|---|---|
| Classes | LSTM | PSO with LSTM | HGO with LSTM | Human group based PSO with LSTM |
| Barren land | 94.50 | 96.92 | 97.80 | 100 |
| Trees | 93 | 97.84 | 98.74 | 99.98 |
| Grasslands | 94.44 | 95.89 | 97.90 | 99.99 |
| Others | 93.92 | 97.78 | 98.89 | 100 |
| Overall | 93.965 | 97.10 | 98.33 | 99.99 |

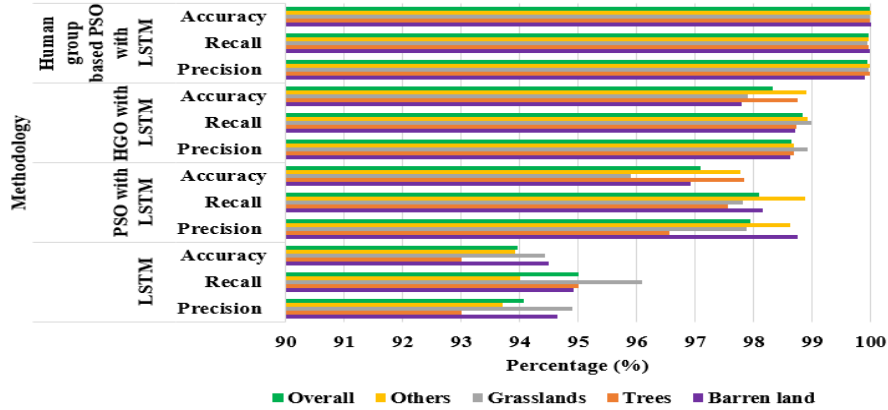

**Figure 8.** Graphical investigation of the proposed model with different optimization techniques on Sat 4 dataset

*4.1.1. Case Study of Proposed LSTM Method with Existing Techniques On Sat 4 Database*

In this subsection, AlexNet [16], ConvNet [16] and VGG [16] are selected for implementing the PSO and HGO techniques. The reason for choosing these techniques is that they are the most widely used neural network architectures for the classification of land use and land cover on satellite images. The selected existing techniques are implemented with HGO and PSO after the max pooling layer of each existing technique. For instance, ConvNet has the output of 4096 features in the max pooling layer and implemented the PSO and HGO as filtering technique that provides only 1000 features as output of fully ConvNet.

**Table 6.** Comparative Analysis of different Neural Networks with PSO and HGO

| Methodology | Dataset | Overall Accuracy (%) |
|---|---|---|
| AlexNet + PSO | Sat 4 | 99.95±0.02 |
| AlexNet + HGO | Sat 4 | 99.94±0.02 |
| AlexNet + PSO + HGO | Sat 4 | 99.96±0.02 |
| ConvNet + PSO | Sat 4 | 99.94±0.02 |
| ConvNet + HGO | Sat 4 | 99.95±0.01 |
| ConvNet + PSO + HGO | Sat 4 | 99.96±0.02 |
| VGGNet + PSO | Sat 4 | 99.95±0.03 |
| VGGNet + HGO | Sat 4 | 99.95±0.03 |
| VGGNet + PSO + HGO | Sat 4 | 99.96±0.02 |
| Proposed LSTM + PSO | Sat 4 | 99.97±0.02 |
| Proposed LSTM + HGO | Sat 4 | 99.97±0.02 |
| Proposed LSTM+ PSO + HGO | Sat 4 | 99.98±0.01 |

Table 6 shows the validated results of proposed LSTM, AlexNet, ConvNet and VGG with PSO and HGO on Sat 4 database in terms of overall accuracy. From the analysis, the results stated that the neural networks and proposed LSTM with PSO and HGO achieved better performance in terms of accuracy on Sat 4 dataset.

*4.2. Quantitative investigation on Sat 6 database*

Sat 6 database is used to evaluate the performance of the proposed model to classify six land cover classes: grassland, water bodies, buildings, barren land, roads and trees. In this case, the performance analysis is carried out by using 40,500 satellite images with 70% of them for training and 30% for testing. The performance of the proposed model is analyzed with different classification techniques DNN, MSVM, LSTM and optimization techniques PSO, HGO and human group based PSO in terms of classification accuracy, recall and precision. The tables 7 and 8 point out that the LSTM classifier achieves better performance in land use and land cover classification, if compared to other classification techniques. The tables 9 and 10 present the performance analysis of the proposed model with different optimization techniques applied to various classes from Sat 6 database. Table 11 describes the case study of LSTM, AlexNet, VGGNet and ConvNet with PSO and HGO on whole Sat 6 dataset in terms of overall classification accuracy. In this database, LSTM classifier attained 99.94% of precision, 99.97% of recall and 99.99% of accuracy. Performance analysis of the proposed model with different classifiers on Sat 6 dataset is then presented in figure 9.

**Table 7.** Performance investigation of the proposed model with different classifiers on Sat 6 dataset by means of recall and precision

| Classes | Human group based PSO with DNN | | Human group based PSO with MSVM | | Human group based PSO with LSTM | |
|---|---|---|---|---|---|---|
| | Precision (%) | Recall (%) | Precision (%) | Recall (%) | Precision (%) | Recall (%) |
| Barren land | 98.97 | 99.02 | 96.09 | 99.07 | 99.92 | 99.99 |
| Trees | 99 | 98.76 | 99 | 98.90 | 99.98 | 99.98 |
| Grasslands | 99.82 | 99.05 | 98.45 | 99.09 | 99.80 | 99.90 |
| Roads | 99.80 | 98.90 | 99.12 | 98.85 | 100 | 100 |
| Buildings | 99.30 | 97.98 | 98.80 | 98.98 | 100 | 99.98 |
| Water bodies | 99.87 | 99 | 99 | 99.80 | 99.98 | 100 |
| Overall | 99.46 | 98.785 | 98.41 | 99.115 | 99.94 | 99.975 |

**Table 8.** Performance investigation of the proposed model with different classifiers on Sat 6 dataset by means of classification accuracy

| | Classification accuracy (%) | | |
|---|---|---|---|
| Classes | Human group based PSO with DNN | Human group based PSO with MSVM | Human group based PSO with LSTM |
| Barren land | 98.90 | 99 | 100 |
| Trees | 98.72 | 98.96 | 99.99 |
| Grasslands | 99.09 | 96.09 | 99.99 |
| Roads | 98.79 | 95.97 | 100 |
| Buildings | 98.90 | 98.06 | 99.99 |
| Water bodies | 99 | 99.5 | 100 |
| Overall | 98.90 | 97.88 | 99.99 |

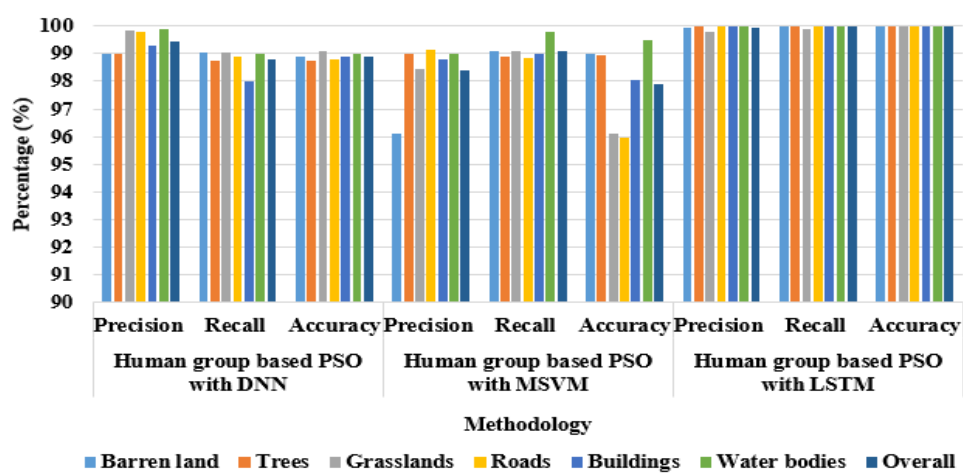

**Figure 9.** Graphical investigation of the proposed model with different classifiers on Sat 6 dataset

**Table 9.** Performance investigation of the proposed model with different optimization techniques on Sat 6 dataset by means of recall and precision

| Classes | LSTM | | PSO with LSTM | | HGO with LSTM | | Human group based PSO with LSTM | |
|---|---|---|---|---|---|---|---|---|
| | Precision (%) | Recall (%) | Precision (%) | Recall (%) | Precision (%) | Recall (%) | Precision (%) | Recall (%) |
| Barren land | 92.90 | 93.09 | 97.78 | 98.08 | 97 | 98.92 | 99.92 | 99.99 |
| Trees | 91.02 | 94.57 | 97.09 | 97.79 | 97.89 | 97 | 99.98 | 99.98 |
| Grasslands | 93.20 | 96 | 98.88 | 97.68 | 96 | 97 | 99.80 | 99.90 |
| Roads | 92.39 | 92.30 | 98.89 | 98.97 | 98.82 | 97.71 | 100 | 100 |
| Buildings | 89 | 94.09 | 97.35 | 98.95 | 97.90 | 98.54 | 100 | 99.98 |
| Water bodies | 92.19 | 93.92 | 97.88 | 98.89 | 98.96 | 98.35 | 99.98 | 100 |
| Overall | 91.78 | 93.995 | 97.97 | 98.39 | 97.61 | 97.87 | 99.94 | 99.975 |

The tables 7 and 8 indicate that the human group based PSO algorithm with LSTM classifier attains better performance in land use and land cover classification in terms of recall, precision and classification accuracy. The proposed model; human group based PSO with LSTM shows maximum of 7.17% and minimum of 2.8% improvement in land use and land cover classification compared to LSTM, PSO with LSTM and HGO with LSTM. The performance investigation of the proposed model with different optimization techniques on Sat 6 database is presented in figure 10.

**Table 10.** Performance investigation of the proposed model with different optimization techniques on Sat 6 dataset by means of classification accuracy

| | Classification accuracy (%) | | | |
|---|---|---|---|---|
| Classes | LSTM | PSO with LSTM | HGO with LSTM | Human group based PSO with LSTM |
| Barren land | 94.09 | 95.52 | 96 | 100 |
| Trees | 92 | 96.73 | 97.08 | 99.99 |
| Grasslands | 94 | 95.90 | 96.85 | 99.99 |
| Roads | 92.03 | 96.57 | 98.80 | 100 |
| Buildings | 91.86 | 95.60 | 96.43 | 99.99 |
| Water bodies | 92.94 | 95.80 | 98 | 100 |
| Overall | 92.82 | 96.02 | 97.19 | 99.99 |

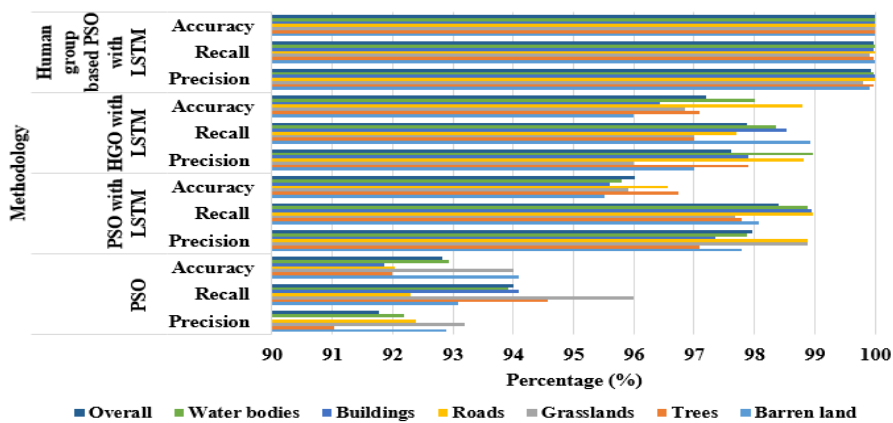

**Figure 10.** Graphical investigation of the proposed model with different optimization techniques on Sat 6 dataset

*4.2.1. Case Study for Proposed LSTM Method with Existing Techniques On Sat 6 Database*

In this subsection, the techniques namely LSTM, AlexNet, VGG and ConvNet are implemented with PSO and HGO on Sat 6 database. For example, eight layers are presented in the AlexNet; first five layers are convolution layers followed by max-pooling layers and the remaining three are fully connected layers. The output of second fully connected layer is 4096 features. Here, the PSO and HGO are implemented to minimize the features. Therefore, the final output of AlexNet is 1000 features. Likewise, the other neural networks are implemented with PSO and HGO after the max-pooling layers. Table 11 presents the validated results on Sat 6 database in terms of overall accuracy.

**Table 11.** Comparative Analysis of Proposed LSTM with existing neural networks on Sat 6 database

| Methodology | Dataset | Overall Accuracy (%) |
|---|---|---|
| AlexNet + PSO | Sat 6 | 99.88±0.03 |
| AlexNet + HGO | Sat 6 | 99.89±0.02 |
| AlexNet + PSO + HGO | Sat 6 | 99.91±0.01 |
| ConvNet + PSO | Sat 6 | 99.86±0.03 |
| ConvNet + HGO | Sat 6 | 99.88±0.02 |
| ConvNet + PSO + HGO | Sat 6 | 99.90±0.01 |
| VGGNet + PSO | Sat 6 | 99.92±0.02 |
| VGGNet + HGO | Sat 6 | 99.92±0.02 |
| VGGNet + PSO + HGO | Sat 6 | 99.93±0.01 |
| Proposed LSTM + PSO | Sat 6 | 99.98±0.02 |
| Proposed LSTM + HGO | Sat 6 | 99.98±0.02 |
| Proposed LSTM+ PSO + HGO | Sat 6 | 99.98±0.01 |

*4.3. Quantitative investigation on Eurosat database*

In this section, Eurosat dataset is used to evaluate the performance of the proposed model to classify 12 land use and land cover classes. Those are permanent crop, annual crop, pastures, river, sea &lake, forest, herbaceous vegetation, industrial building, highway and residential building. In this scenario, the performance analysis is accomplished for 27,000 satellite images with 70% of the data used for training and 30% for testing with two case studies. Initially, Table 12 presents the performance value of the proposed model for different classes of land use and land cover classification in terms of accuracy, recall and precision, with some existing classifiers DNN, MSVM, LSTM and the optimization techniques PSO, HGO and Human Group based PSO. The accuracy, recall and precision of the LSTM classifier with Human Group based PSO is 97.40%, 98.70% and 97.80%, respectively. The LSTM classifier with Human Group based PSO shows an improvement in land use and land cover classification. Performance analysis of the proposed model on Eurosat dataset is presented in figure 11.

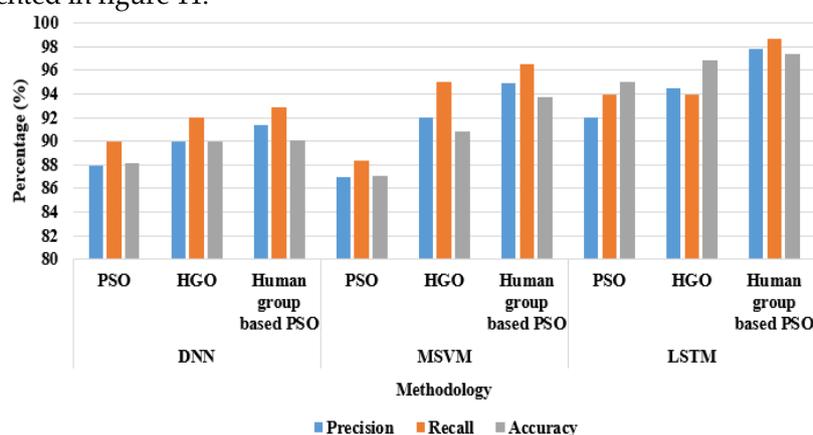

**Figure 11.** Graphical investigation of the proposed model on Eurosat dataset

**Table 12.** Performance investigation of the proposed model on Eurosat dataset

| Classification | Optimization techniques | Average value | | |
| --- | --- | --- | --- | --- |
| | | Precision (%) | Recall (%) | Accuracy (%) |
| DNN | PSO | 87.89 | 90 | 88.2 |
| | HGO | 90 | 92.02 | 90 |
| | HGO + PSO | 91.38 | 92.91 | 90.09 |
| MSVM | PSO | 87 | 88.39 | 87.03 |
| | HGO | 92.04 | 95 | 90.85 |
| | HGO + PSO | 94.90 | 96.50 | 93.70 |
| LSTM | PSO | 92 | 93.98 | 95 |
| | HGO | 94.50 | 94 | 96.90 |
| | HGO + PSO | 97.80 | 98.70 | 97.40 |

*4.3.1 Case Study of Implementing PSO and HGO On Proposed Method*

In this subsection, GoogleNet [15] is implemented with optimization techniques namely PSO and HGO and selected only 1000 features. The other techniques like AlexNet [16], VGG [16] and ConvNet [16] worked only on the Sat 4 and Sat 6 database. Hence, these techniques in [16] are not considered in the case study of Eurosat database. While comparing with AlexNet and ConvNet, the GoogleNet uses techniques namely 1×1 convolutions in the middle of the architecture and global average pooling. In addition, the inception module is also different than other architectures. The 1×1, 3×3, 5×5 convolutions are presented in the inception module and 3×3 max pooling is operated in a parallel way and the input and output are stacked together for generating the final output. Table 13 shows the comparative analysis of LSTM and GoogleNet [15] with PSO and HGO on the whole Eurosat dataset.

**Table 13.** Comparative Analysis of Proposed LSTM with PSO and HGO on Eurosat dataset

| Methodology | Dataset | Overall Accuracy (%) |
| --- | --- | --- |
| GoogleNet + PSO | Eurosat | 96.18±0.3 |
| GoogleNet + HGO | Eurosat | 96.20±0.2 |
| GoogleNet + PSO + HGO | Eurosat | 96.40±0.2 |
| Proposed LSTM + PSO | Eurosat | 97.37±0.03 |
| Proposed LSTM + HGO | Eurosat | 97.39±0.02 |
| Proposed LSTM+ PSO + HGO | Eurosat | 97.40±0.01 |

*4.4. Comparative analysis*

The comparative analysis between the proposed and existing models is represented in table 11. Analyzing recent works from the literature, based on similar data, it was found that Helber et al. [15] developed a new patch based land use and land cover classification technique using Eurosat database. This work explained how CNN was used to detect the land use and land cover changes which helped in improving the geographical maps. Unnikrishnan et al. [16] implemented a novel deep learning method for three different networks VGG, AlexNet and ConvNet, and Sat 4 and Sat 6 datasets were used to analyze the performance of the developed model. Papadomanolaki et al. [17] designed a deep learning model based on a CNN for accurate land cover classification, by including 2 band information (red and near infrared) with a reduced number of filters, which were tested and trained to classify the images into different classes. The model proposed in this manuscript was compared with earlier models in terms of precision, recall and accuracy.

The table 14 illustrates that the proposed model achieves a minimum of 0.01% and a maximum of 2.56% improvement in accuracy on Sat 4, Sat 6 and Eurosat datasets. In this study, human group based PSO algorithm is combined with LSTM classifier in order to gain better performance in land use and land cover classification. The proposed Human Group based PSO algorithm significantly reduces the "curse of dimensionality" issue and this helps the LSTM classifier to achieve a better performance in the classification.

Table 14. Comparative investigation of the proposed and existing models

| Methodology | Dataset | Overall Accuracy (%) |
|---|---|---|
| GoogleNet [15] | Eurosat | 96.69 |
| 2 band AlexNet [16] | Sat 4 | 99.66 |
|  | Sat 6 | 99.08 |
| Hyper parameter tuned AlexNet [16] | Sat 4 | 98.45 |
|  | Sat 6 | 97.43 |
| 2 band ConvNet [16] | Sat 4 | 99.03 |
|  | Sat 6 | 99.10 |
| Hyper parameter tuned ConvNet [16] | Sat 4 | 98.45 |
|  | Sat 6 | 97.48 |
| 2 band VGG [16] | Sat 4 | 99.03 |
|  | Sat 6 | 99.15 |
| Hyper parameter tuned VGG [16] | Sat 4 | 98.59 |
|  | Sat 6 | 97.95 |
| AlexNet [17] | Sat 4 | 99.98 |
|  | Sat 6 | 99.92 |
| ConvNet [17] | Sat 4 | 99.98 |
|  | Sat 6 | 99.90 |
| VGG [17] | Sat 4 | 99.98 |
|  | Sat 6 | 99.96 |
| Proposed model | Eurosat | 97.40 |
|  | Sat 4 | 99.99 |
|  | Sat 6 | 99.99 |

## 6. Conclusions

The objective of this study was to propose an effective feature selection model to classify the land use and land cover classes of urban and agricultural environment. The proposed model helped to analyze the changes in land productivity, soil-quality, biodiversity, for instance, that provided a clear idea about environmental quality, wildlife habitat, human habitat, loss of prime agricultural lands, uncontrolled development, etc. In this study, an optimization procedure based on the combination of LGBPHS, HOG and Haralick texture features was first utilized to extract the feature vectors of the objects from the normalized remote sensing images. The Human Group based PSO algorithm was then applied to select the optimal feature vectors that helped in further improving the performance of classification.  The optimal selected features were given as the input to a LSTM classifier. The proposed model achieved a better performance when compared to the existing models in land use and land cover classification in terms of recall, accuracy and precision. The simulation result showed that the proposed model achieved a maximum of 2.56% enhancement in classification accuracy on Sat 4, Sat 6 and Eurosat databases. In addition, the computational time of the proposed model is 1.24 seconds. In the future work, an optimization based clustering approach will be included in the proposed model to verify if the land use and land cover classification method can be further improved.